\documentstyle{article}
\title{\LARGE Kinematic in $SU(2,2)\simeq O(2,4)$ alternative world (observable coordinates and impulses representation)}
\author{A.~N.~Leznov\thanks{ Universidad Autonoma del Estado de Morelos, CCICAp,Cuernavaca, Mexico}} \date{}

\newcommand{\rig}[2]{\stackrel{#2\rightarrow}{#1}}

\begin{document}
\maketitle

\maketitle

\begin{abstract}

The matrix elements connected non physic angles space of representation with the space of physic observable of alternative $O(2,4)$ world found on explicit form. All such matrix elements are represented in terms of solutions of Gauss hypergeometric equations.     
From these results wave function of free moving particle in alternative world may be constructed. Definite relations between micro and macro label variables are presented.

\end{abstract}

\section{Introduction}

Something about 150 years ago Lobachevski, Gauss and Bolyai have doubted in a fact that Geometry of the world on big distances is the Euclidean. They independently have found a positive answer to the question of how this may be happen by discovering the spaces of constant curvatures. During the last 150 years it became clear that Geometry by itself is the consequence of fundamental physical laws of the nature. And thus question of these scientists may reformulate as follows -How the laws of nature may be changed at big distances ( bug energy or other observable), where up to now they have no experimental conformation? 
From the time of Newton our world (after Einstein modification) describes in terms of 4 coordinates $x_i$, 4 $p_i$-impulses, $F_{ij}$- 6 generators of Lorenz transformation and I-unity element. And thus the main kinematic variables of physics are distances, impulses and angular moments ($I$ is dimensionless). By this reason it is natural to wait the arising the constants of the same dimensions in the main laws of physics. The extremal will be situations when physics values in experiment are the same order of these constants. 
Below we present the most general form of the commutation relations between the elements of the alternative four-dimensional space-time and its group of motion ($x_i$-coordinates, $p_i$-impulses, $F_{ij}$-generators of Lorenz transformation, I-unity element) This commutation relations are the following ones \cite{KHL} (they are written under assuming that Lorenz group is included into the main laws of the nature) 
$$
[p_i,x_j]= ih(g_{ij}I+{F_{ij}\over H}),\quad [p_i,p_j]= {ih\over L^2}F_{ij},\quad [x_i,x_j]= {ih\over M^2}F_{ij}
$$
\begin{equation}
[I,p_i]= ih(-{p_i\over H}+{x_i\over L^2}),\quad [I,x_i]= ih(-{p_i\over M^2} +{x_i\over H}),\quad[I,F_{ij}]=0 \label{1}
\end{equation} 
$$
[F_{ij},x_s]= ih(g_{is}x_j-g_{js}x_i),\quad [F_{ij},p_s]= ih(g_{is}p_j-g_{js}p_i)
$$
$$
[F_{ij},F_{sk}]= ih(g_{js}F_{ik}-g_{is}F_{jk}-g_{jk}F_{is}+g_{ik}F_{js})
$$ 
(Commutation relations of Minkowski space-time arises from (\ref{1}) under infinite values of dimensional parameters the length $L$, the impulse $M\to Mc$ and the action $H$). Commutation relations (\ref{1}) must be supported by some additional conditions which responsible for correct limit to usual Minkowski space-time in the infinite limit of dimensional parameters. Such conditions looks as 
$$
IF_{i,j}={x_j p_i -x_i p_j+p_i x_j-p_j x_i\over 2}
$$ 
which in Minkowski limit ($I\to 1$) represent relation between angular moments and linear coordinates and impulses. The additional conditions above allow to choose definite representation of algebra (\ref{1}) in the unique way \cite{LEZM},\cite{LEZp}. 
The position of Plank constant $h$ in (\ref{1}) separate all physics on quantum and classical domains. In the case of consideration problems in classical physics in (\ref{1}) it is necessary to perform in all formulae above limiting procedure 
$$
{[A,B]\over ih}\to \{A,B\}
$$
(change all commutator of quantum theory on corresponding Poisson brackets) and consider (\ref{1}) on the level of functional group \cite{ESEN}, (see for instance \cite{LM}).

The equalities of Jacobi are satisfied for (\ref{1}). It should be stressed the signs of 
$L^2,M^2$ are not required to be positive. 
Particular cases of (\ref{1}) have been arises before in so called quantum spaces. The main idea of which was changing properties of Minkowski space-time only on micro scale. 
The limiting procedure $M^2,H\to \infty$ leads to the space of constant curvature (LGB), considered in connection with Column problem by E.Schredinger \cite{SCHR}, $L^2,H\to \infty $ leads to quantum space of Snyder \cite{SNY}, $H\to \infty$ leads to Yangs quantum space \cite{Y}. Except of $L^2,M^2$ parameter dimension of action $H$ was introduced in \cite{KHL}. 

In general (\ref{1}) is commutation relations one of real forms of six-dimensional rotation group $O(1,5),O(2,4),O(3,3)$. As was mentioned above realization of algebra (\ref{1}) with necessary additional condition was found before \cite{LEZM} in four dimensional representation space of two unit vectors $O(1),O(5);O(2),O(4);O(3),O(3)$ or in four dimensional un physics Minkowski space \cite{LEZp}. Of course these two representation are equivalent one. From mathematical point of vie the first one is more natural (these are rotation groups). From the point of vie of physics interpretation more attractive is the second one. Relations between these two representations reader may find in Appendix I. We emphasize that volume of the space of representation is finite one.
 
But to find realization is only half of the deal. The space of representation in both cases are not physical ones. And to have a physical theory it is necessary to have matrix elements of transition from representation space to the same one in the space with physically observable. 
In the usual theory the most used are $x$ representation space and $p$ one. In the first one case all functions depend on 4 space-time coordinates, which can be mesuared simultaneously. In alternative space situation is more complicate. It is only possible to find proper values of 4 mutually commutative operators and by these values enumerate basis functions of $x$representation of the real alternative world. The same is true with respect to $p$ representation space. 

In the usual Minkowski space interval is invariant with respect to Lorenz transformations and commutative with all coordinates. The same property satisfy the Kazimir operator of the second order of the algebras $O(1,4),O(2,3)$ $K_2=x_4^2-\rig{x}{}^2+{\rig{f}{}^2-\rig{l}{}^2\over M^2}$.
Thus in alternative space the last expression will be considered as interval operator. 
Two additional quantum numbers may be connected with $\rig{l}{}^2,l_z$, or two angles by which fixed direction in the space. The last remaining number may be connected  
with proper value of time operator or $\rig{x}{}^2$ . Absolutely the same is situation in real alternative $p$ space. In the usual theory the mass is invariant to all 10 transformation of Poincare group and thus in alternative space naturally define 
$m^2=K_2=p_4^2-\rig{p}{}^2+{\rig{f}{}^2-\rig{l}{}^2\over L^2}$, which is invariant to 10 transformation of de-Sitter group. 
All additional quantum number may be chosen by the same way as it was done for $x$ representation above. 

The goal of the present paper partially realize above program and construct wave functions of free motion in explicit form. We will be able to construct in terms of solution of hypergeometrival Gauss equation matrix elements of transition from unphysical basis to the physical one in x and p representation. But investigate properties oh the matrix elements between physical states of alternative world up to now was not possible for us.

The second order Kazimir operator in terms of physical observable variables (after performing to diagonal form) looks as 
\begin{equation}
K_2(O(2,4)) = I^2 - {(x - {L^2\over H}p)^2\over L^2} +( {h^2\over H^2} - {h^2\over L^2M^2} )( {L^2\over h^2}p^2 + {\rig{f}{}^2-\rig{l}{}^2\over h^2}) \label{KAZ} 
\end{equation}
In what follows we use notations 
$$
\delta^2=({h^2\over H^2} - {h^2\over L^2M^2}), \quad \bar x=(x - {L^2\over H}p) 
$$
Irreducible representation of non compact algebras $O(p,q)$ (for representation under consideration) is defined by arbitrary complex number $\rho$. Second Kazimir operator (quadratic on generators with positive sign before non compact terms )  for these representation equal $K_2(O(p,q)) = \rho(\rho+(p+q-2))$. Three possibility for unitary representations are the following ones $\rho=-{p+q-2\over 2}+id$-the main continuous series, $-{p+q-2\over 2}\leq \rho \leq {p+q-2\over 2}$-additional series,$\rho=k$-natural number -discrete series. 

\section{Realization the alternative $O(2,4)$ space with $O(2,3)$ group of motion}

This case arises under condition $L^2 \leq 0 \leq \delta^2$ in (\ref{KAZ}).

In this case we will use angular representation. All formulae for generators of alternative space are written in terms of 4 angles of four dimensional unite vector $q^1=( =(\sin \tau \sin \theta \sin \psi,\sin \tau \sin \theta \cos \psi, \sin \tau \cos \theta,\cos \tau)$ and two dimensional one $p^1 =(\sin \phi,\cos \phi)$. 

In the case under consideration \cite{LEZM} 7 generators are compact ones  
$$
{\delta\over h}l_{\alpha}={i\over 2}\sum \epsilon _{\alpha,\beta,\gamma}Q_{\beta,\gamma},\quad {\delta L\over h}p_4=i\frac{\partial }{\partial \phi}
\quad {\bar x_{\alpha}\over L}=i Q_{\alpha,4}
$$
are compact one and 8 are non compact 
$$
-iI=\cos \phi(-\rho q^1_4+ \sum q^i_4 Q^{i,1})+
\sin \phi q^1_4 \frac{\partial}{\partial \phi}
$$
$$
-i{\bar x_4\over L}=-\sin \phi(-\rho q^1_4+ \sum q^i_4 Q^{i,1})+
\cos \phi q^1_4 \frac{\partial}{\partial \phi}
$$
$$
-i{\delta L\over h} p_{\alpha}=\cos \phi(-\rho q^1_{\alpha}+ \sum q^i_{\alpha} Q^{i,1})+
\sin \phi q^1_{\alpha} \frac{\partial}{\partial \phi}
$$
\begin{equation}
-i{\delta \over h} f_{\alpha}=-\sin \phi(-\rho q^1_{\alpha}+ \sum q^i_{\alpha} Q^{i,1})+
\cos \phi q^1_{\alpha} \frac{\partial}{\partial \phi} \label{REP}
\end{equation}
where $Q_{ij},Q^{ij},1\leq i,j \leq 4$ generators if left (right) translations of compact four dimensional group of rotation. We pay attention the reader that all formulae above do not contain Plank constant- $\delta$ is proportional to it.

\subsection{Impulse representation} 

In connection of its definition 
$$
m^2=p_4^2-\rig{p}{}^2-{\rig{f}{}^2-\rig{l}{}^2\over L^2}
$$
In formula above and in what follows $0\leq  L^2$. 
Using explicit form of generators (\ref{REP}) we obtain (${L^2\delta^2\over h^2}m^2\to
m^2$)
$$
m^2=\cos^{\rho} \tau (\sum_{i,k,\alpha} q^i_{\alpha} Q^{i,1} q^k_{\alpha} Q^{k,1})+\sum_k q^1_4 q^k_4 Q^{k,1}-(q^1_4)^2)\frac{\partial^2}{\partial \phi^2})\cos^{-\rho}\tau=
$$
$$
\cos^{\rho}\tau(\cos^2\tau \frac{\partial^2}{\partial \tau^2}+2{\cos\tau\over \sin\tau}\frac{\partial}{\partial \tau}+{\cos^2\tau\over \sin^2\tau}\rig{l}{}^2-\cos^2\tau \frac{\partial^2}{\partial \phi^2})\cos^{-\rho}\tau 
$$  
Now problem is in resolving on proper values and functions equation
\begin{equation}
\cos^2\tau \frac{\partial^2}{\partial \tau^2}+2{\cos\tau\over \sin\tau}\frac{\partial}{\partial \tau}+{\cos^2\tau\over \sin^2\tau}\rig{l}{}^2-\cos^2\tau \frac{\partial^2}{\partial \phi^2})\tilde F=k(k+3)\tilde F \label{BUIT}
\end{equation}
where $m^2=k(k+3)$ and $\rig{l}{}^2=\frac{\partial^2}{\partial \theta^2}+{\cos\theta\over \sin\theta}\frac{\partial}{\partial \theta}+{1\over \sin^2\theta}\frac{\partial^2}{\partial \psi^2}$, $\tilde F=\cos^{-\rho}\tau F $.
In what follows we will work in the basis with diagonal operator $\rig{l}{}^2=-l(l+1)$.
Thus it will be necessary only one operator commutative with $\rig{l}{}^2,l_3$ and $m^2$ to construct basis in $p$ representation.

\subsubsection{Energy basis}

In this case basis functions are enumerated by proper values of four mutually commutative operators $\rig{l}{}^2,l_3,m^2,{\delta L\over h}p_4\equiv \epsilon=i\frac{\partial}{\partial \phi}$. ($\epsilon$ arbitrary natural number). The main equation (\ref{BUIT}) takes form of ordinary differential equation of the second order
$$
\frac{d^2 F}{d\tau^2}+{2\over \cos\tau \sin\tau}\frac{d F}{d \tau}-({l(l+1)\over \sin^2\tau}-\epsilon^2-{k(k+3)\over \cos^2\tau})F=0
$$ 
By the exchange of argument $\cos^2\tau \to z$ the last equation pass to
$$
4z(1-z)F_{zz}-2(2z+1)F_z-({l(l+1)\over 1-z}-\epsilon^2+{k(k+3)\over z})F=0
$$
By exchange of the function $F\to z^{\alpha}(1-z)^{\beta}F$ with $2\alpha=k+3,(-k),2\beta=l,-(l+1)$ the last equation takes form of Gauss equation for hypergeometric function
with $a={k+3+l+\epsilon\over 2},b={k+3+l-\epsilon\over 2},c=k+{5\over 2}$. Finally for matrix
of transition from angles state of representation to alternative space with quantum numbers
$m^2=({h\over \delta L})^2 k(k+3), p_4={h\over \delta L}\epsilon, l,l_3$ we obtain 
\begin{equation}
\cos^{\rho}\tau\cos^{k+3}\tau \sin^l\tau F({k+3+l+\epsilon\over 2},{k+3+l-\epsilon\over 2},k+{5\over 2},\cos^2\tau)e^{i\phi \epsilon}Y^l_{l_3}(\rig{n}{})\label{E}
\end{equation}
where $Y^l_{l_3}(\rig{n}{})$ - usual spherical harmonic function.

Condition that matrix element have no singularities lead to quantization $-n={k+3+l-\epsilon\over 2}$ from what follow that $k$ is natural non negative number. Except of this we obtain relation between energy, mass, orbital momentum and radial impulses, as it possible interpreted new quantum number $n$ (?) $\epsilon=k+3+l+2n$

\subsubsection{ $\rig{p}{}^2$ basis}

Four mutually commutative operators $\rig{l}{}^2,l_3,m^2,\rig{p}{}^2$ define basis in that case.
In connection with (\ref{REP}) we have for dimensionless (we omitted below tangent transformation connected with factor $\cos^{\rho} \tau$) 
$$
-\rig{p}{}^2=\cos^2\phi((\cos^2\tau \frac{\partial}{\partial \tau})^2+2{\cos\tau\over \sin\tau}\frac{\partial}{\partial \tau}+{1\over \sin^2\tau}\rig{l}{}^2)+\sin^2\tau(\sin\phi\frac{\partial}{\partial \phi})^2-
$$
$$
2\cos\tau \sin\tau\frac{\partial}{\partial \tau}\cos\phi \sin\phi\frac{\partial}{\partial \phi}+(-3+\sin^2\tau)\cos\phi \sin\phi\frac{\partial}{\partial \phi}+\sin^2\phi\cos\tau \sin\tau\frac{\partial}{\partial \tau}
$$
All terms with derivatives may be regrouped to very simple form
$$
(\cos\tau \cos\phi\frac{\partial}{\partial \tau}-\sin\tau \sin\phi\frac{\partial}{\partial \phi})+2{\cos\phi\over \sin\tau}(\cos\tau \cos\phi\frac{\partial}{\partial \tau}-\sin\tau \sin\phi\frac{\partial}{\partial \phi})
$$
from which it follows that operator $\rig{p}{}^2$ is commutative with arbitrary function of argument ${\cos\phi\over \sin\tau}$ and equations on proper values and functions for commutative operators $m^2,\rig{p}{}^2$ are divided in variables $u={\sin\phi\over  \cos\tau},v={\cos\phi\over \sin\tau}$. Equation on proper value of operator $\rig{p}{}^2$ looks as
$$
(1-v^2)^2V_{vv}+(\rig{p}{}^2-l(l+1)v^2)V=0
$$
and (\ref{BUIT}) takes the form
$$
(u^2-1)U_{uu}+4uU_u+({\rig{p}{}^2-l(l+1)\over u^2-1}-k(k+3))U=0
$$
By changing the arguments $x=v^2,y=u^2$ transform them to the form
\begin{equation}
4x(1-x)V_{xx}+2(1-x)V_x+({\rig{p}{}^2-l(l+1)\over 1-x}+l(l+1))V=0 \label{V}
\end{equation}
\begin{equation}
4y(1-y)U_{yy}+(2-10y)U_y+{\rig{p}{}^2-l(l+1)\over 1-y}+k(k+3))U=0 \label{U}
\end{equation}
Now it is necessary to take into account limitations on $v^2,u^2$ which follows from their definition. Equations which define $u,v$ functions resolved in back direction lead to
$$
\cos^2\phi={1-v^2\over u^2-v^2},\quad cos^2\tau={u^2-1\over u^2-v^2}
$$
and thus the following limitation on $v^2,u^2$ take place $0\leq v^2 \leq  1,1\leq u^2$ or visa versa. 

In the first case $0\leq x \leq 1,1\leq y, 0\leq {1\over y} \leq 1 $.     
And thus to have the same interval for both variables it is necessary in (\ref{U}) change variable on inverse one. After such transformation it takes the form
\begin{equation}
4y(1-y)U_{yy}-2(1+3y)U_y+{\rig{p}{}^2-l(l+1)\over 1-x}-{k(k+3)\over y})U=0 \label{U1}
\end{equation}
and after transformation $U\to (1-y)^{\beta}y^{\alpha} U$ with $2\alpha=k+3,2\beta(2\beta+2)+
\rig{p}{}^2-l(l+1)=0$, it pass to Gauss equation with $a=\alpha+\beta,b=\alpha+\beta+{1\over 2},c=2\alpha-{1\over 2}$.
The equation (\ref{V})after transformation $V\to (1-x)^{\alpha} V$  both equations pass to equations of hypergeometric Gauss function $F(a,b,c,x)$
In the first case $ 2\alpha(2\alpha-2)+\rig{p}{}^2-l(l+1)=0$, $a=\alpha+{l\over 2},b=\alpha-{l+1\over 2},c={1\over 2}$. Condition that on considered interval hypergeometric function
has no  singularities are following ones

In the second case $0\leq y \leq 1,1\leq x, 0\leq {1\over x} \leq 1 $. It is change on inverse argument in (\ref{V}). After such transformation it takes the form
\begin{equation}
4x(1-x)V_{xx}+6(1-x))V_x+{\rig{p}{}^2-l(l+1)\over 1-x}-{l(l+1)\over x})V=0 \label{V1}
\end{equation}
and $,U\to (1-y)^{\beta} U$In the second case $\beta=\alpha-1$,$a=\alpha+{k+1\over 2},b=\alpha-{k+2\over 2},c={1\over 2}$,$z=y$. The wave function of transition looks in this case as
\begin{equation}
(1-v^2)^{\alpha}(1-u^2)^{\alpha-1}F(\alpha+{l\over 2},\alpha-{l+1\over 2},{1\over 2},v^2)
F(\alpha+{k+1\over 2},\alpha-{k+2\over 2},{1\over 2},u^2)Y^l_{l_3}(\rig{n}{})\label{p^2}
\end{equation}
where $Y^l_{l_3}(\rig{n}{})$ - usual spherical harmonic function. 

Resolving of these limitation is as follows
$$ 
v^2=\cos^2\theta_1,\quad {1\over u^2}=\cos^2\theta_2,\quad \cos^2\phi={\cos^2\theta_1\sin^2\theta_2\over 1-\cos^2\theta_1\cos^2\theta_2},\quad \cos^2\tau={\cos^2\theta_2\sin^2\theta_1\over 1-\cos^2\theta_1\cos^2\theta_2}
$$
The equation with respect to $u^2$ (\ref{U}) must be transformed to inverse value of its argument $u^2\to {1\over u^2}$. After such transformation it takes the form
$$
4y(1-y)U_{yy}-2(1+3y)U_y+{\rig{p}{}^2-l(l+1)\over 1-x}-{k(k+3)\over y})U=0 
$$
and after transformation $U\to (1-y)^{\beta}y^{\alpha} U$ with $2\alpha=k+3,2\beta(2\beta+2)+
\rig{p}{}^2-l(l+1)=0$, ot pass to Gauss equation with $a=\alpha+\beta,b=\alpha+\beta,+{1\over 2},c=2\alpha-{1\over 2}$.

\subsubsection{ $\rig{f}{}^2$ basis}

In this basis together with $m^2,\rig{l}{}^2,\rig{f}{}^2$ diagonal also operator $p_4^2-\rig{p}{}^2$, which is the square of mass in Minkowski space-time.
Directly from (\ref{REP}) it easy to see that generator $\rig{f}{}$ may be obtain from $\rig{p}{}$ by simple exchange $\cos\phi\to- \sin\phi,\sin\phi\to \cos\phi$ and thus the same substitution it is necessary to do in all formulae of the previous section.

\subsection{Relation between the energy and square of impulse}

As was noticed above in alternative world it is not possible to measure energy and square of impulse simultaneously (generators of them are not commutative). But it is only possible to find distribution of squares of impulses in the state with given energy or visa versa. The measure of this is  the matrix element between the states with fixed energy and fixed square of impulses. This will be answer on the question what is the part of kinetic energy in the total one in probability sense.  
Matrix element of such transition is constructed from solutions of (\ref{p^2}) and (\ref{E})
$$
\int d V(angles) F(m^2,p^2,l^2,l_3,angles)\bar F(m^2,\epsilon,l^2,l_3,angles)   
$$
In the domain of non extremely values of $m^2,p^2,\epsilon$ it must have maximum around
classical relation $m^2+p^2=\epsilon^2$. At the present time the way of calculation of this  
integral or investigation it in extremal conditions is unknown to the author. Corresponding calculation in the case $O(2,2)$ alternative space reader can find in Appendix II.  

\subsection{Coordinate representation} 

In the case under consideration the symmetry of $m^2$ operator is defined by $O(2,3)$ algebra.
But symmetry of interval operator $x^2$ is defined by the sign of $M^2$ constant. Thus twice
possibilities are possible. 

From (\ref{REP}) it follows
$$
{x_{\alpha}\over L}=i(Q_{\alpha,4}-{h\over \delta H}(Q_{\alpha,6}),\quad {x_4\over L}=i(Q_{5,4}-{h\over \delta H}(Q_{5,6})
$$
Introducing Lorenz index $i=5,\alpha$ in $Q_{i,6}$ we rewrite relation above in equivalent form
$$
e^{Q_{6,4}\sigma}{x_i\over L}e^{-Q_{6,4}\sigma}=ie^{Q_{6,4}\sigma}(Q_{i,4}-{h\over \delta H}(Q_{i,6})e^{-Q_{6,4}\sigma}=
$$
$$
i((\cosh\sigma-{h\over \delta H}\sinh\sigma)Q_{i,4}+(\sinh\sigma-{h\over \delta H}\cosh\sigma)Q_{i,6})
$$
In writing of the last equalities we took into account that Lorenz index $i$ is commutative with operator $Q_{6,4}$ and the fact that indexes $4,6$ are Lorenz vector with respect to the last transformation.

Two possibilities take place ${h\over \delta H}\leq 1$ or $1 \leq {h\over \delta H}$.
In the first one choosing $\tanh \sigma={h\over \delta H}$ and having in a consequence
${1\over \cosh^2\sigma}=-{h^2\over \delta^2M^2L^2}\equiv -({h\over \delta M L})^2 $ we conclude that last relation possible only under additional condition $0\leq M^2$ ($L^2\leq 0$). In this case $x_i$ after simple manipulation may be presented in a form
$$
x_i=i{h\over \delta M}e^{-Q_{6,4}\sigma}Q_{i,4}e^{Q_{6,4}\sigma}
$$
and interval operator $(Int)^2$ (see Introduction) looks as
\begin{equation}
(Int)^2=x_4^2-\rig{x}{}^2+{\rig{f}{}^2-\rig{l}{}^2\over M^2}=({h\over \delta M})^2e^{-Q_{6,4}\sigma}((Q^2_{\alpha,4}-Q^2_{\alpha,5}-Q^2_{5,4}+Q^2_{\alpha,\beta}))e^{Q_{6,4}\sigma}
\label{IP}
\end{equation}
and invariant with respect to $O(1,4)$ transformations of $x,F$ algebra. 

In the second case ($1 \leq {h\over \delta H}$ it possible choose $(\tanh \sigma)^{-1}={h\over \delta H}$ having in a consequence ${1\over \sinh^2\sigma}={h^2\over \delta^2M^2L^2}\equiv ({h\over \delta M L})^2 $ we conclude that last relation possible only under additional condition $M^2\leq 0$ ($L^2\leq 0$). 
On this case
$$
x_i=i{h\over \delta M}e^{-Q_{6,4}\sigma}Q_{i,6}e^{Q_{6,4}\sigma}
$$
and interval operator looks as
\begin{equation}
(Int)^2=x_4^2-\rig{x}{}^2-{\rig{f}{}^2-\rig{l}{}^2\over M^2}=({h\over \delta M})^2e^{-Q_{6,4}\sigma}((Q^2_{\alpha,6}+Q^2_{\alpha,5}-Q^2_{6,5}-Q^2_{\alpha,\beta}))e^{Q_{6,4}\sigma}
\label{IN}
\end{equation}
and invariant with respect to $O(2,3)$ transformations of $x,F$ algebra. 
Quadratic on operators of $O(2,4)$ algebra in (\ref{IN}) was been calculated before (\ref{BUIT}). By the same way operator in (\ref{IP}) is calculated and lead to (\ref{IPP}).  

Below we present equations on the proper values and functions in these two cases in dimensionless form (in calculation below squares of compact generators have positive sign)
\begin{equation}
(\cos^2\phi K_2(O(4))-2\cos\phi\sin\phi\frac{\partial}{\partial \phi}-\cos^2\phi\frac{\partial^2}{\partial \phi^2})\tilde F=-t(t+3)\tilde F \label{INN}
\end{equation}
where $\tilde F=\cos^{-\rho}\phi e^{Q_{6,4}\sigma} F$ and $(Int)^2=t(t+3)({h\over \delta M})^2$ in the case $0\leq M^2$.
\begin{equation}
(\cos^2\tau \frac{\partial^2}{\partial \tau^2}+2{\cos\tau\over \sin\tau}\frac{\partial}{\partial \tau}+{\cos^2\tau\over \sin^2\tau}\rig{l}{}^2-\cos^2\tau \frac{\partial^2}{\partial \phi^2})\tilde F=t(t+3)\tilde F \label{IPP}
\end{equation}
where $\tilde F=\cos^{-\rho}\tau e^{Q_{6,4}\sigma} F$ and $(Int)^2=t(t+3)({h\over \delta M})^2$ in the case $M^2\leq 0$.
In both cases it is necessary to know action of translation operator $e^{Q_{6,4}\sigma}$ on function from representation space. 

\subsubsection{Action of operator $e^{Q_{6,4}\sigma}$} 

From (\ref{REP}) we have explicit expression for anti hermitian operator $Q_{6,4}$
$$
Q_{6,4}=-\rho\cos\tau \cos\phi+\cos\phi \sin\tau \frac{\partial}{\partial \tau} +\cos\tau \sin\phi \frac{\partial}{\partial \phi}\equiv -\rho\cos\tau \cos\phi+A  
$$
which symmetrical with respect to permutation of $\phi ,\tau$ variables and operator $A$ contain terms with derivatives. The last expression  can be represented in two equivalent forms 
$$
Q_{6,4}=\cos^{\rho}\tau (-\rho{\cos\phi\over \cos\tau} +A)\cos^{-\rho}\tau =
\cos^{\rho}\phi(-\rho{\cos\tau\over \cos\phi} +A)\cos^{-\rho}\phi 
$$
Up to tangent transformation calculation of $e^{Q_{6,4}\sigma}$ is equivalent to calculation
$g\equiv e^{(-\rho{\cos\tau\over \cos\phi} +A)\sigma}$

We have following commutation relations
$$
[A,{\cos\phi\over \cos\tau}]=(({\cos\phi\over \cos\tau})^2-1),\quad
[A,{\cos\tau\over \cos\phi}]=(({\cos\tau\over \cos\phi})^2-1)
$$
Further
$$
\frac{d g}{d \sigma}\equiv \dot g=(-\rho{\cos\tau\over \cos\phi} +A)g
$$
Let us find solution of this equation on the form $g=e^{A\sigma}s$
Substituting this ansatz into equation for $g$ we obtain equation for $s$
$$ 
\dot s=-\rho e^{-\sigma A}{\cos\tau\over \cos\phi}e^{\sigma A} s\equiv \rho c s
$$
Coefficient on equation for $c$ is usual function of argument ${\cos\tau\over \cos\phi}$.
The most simple why to calculate it is to consider equation it satisfied.
$$
\dot c=(c^2-1),\quad {c-1\over c+1}=c_0 e^{2\sigma},\quad c={1-c_0e^{2\sigma}\over 1+ c_0e^{2\sigma}}
$$
with initial condition $c_0={{\cos\phi\over \cos\tau}-1\over {\cos\phi\over \cos\tau}+1}$
Now equation for $s$ is resolving with result (initial condition is taken into account)
$$
s=({1+c_0\over e^{-\sigma}+ c_0e^{\sigma}})^{\rho}=(\cosh\sigma-{\cos\tau\over \cos\phi}\sinh\sigma)^{-\rho}
$$
Gathering all results above we obtain for operator $e^{Q_{6,4}\sigma}$
$$
e^{Q_{6,4}\sigma}=\cos^{\rho}\tau e^{\sigma A}(\cosh\sigma-{\cos\tau\over \cos\phi}\sinh\sigma)^{-\rho}\cos^{-\rho}\tau
$$
Action of translation operator $e^{\sigma A} F(\tau,\phi)=F(\tilde \tau,\tilde \phi)$ with
$$
\tan {\tilde \tau-\tilde \phi\over 2}=e^{\sigma}\tan {\tau-\phi\over 2},\quad
\tan {\tilde \tau+\tilde \phi\over 2}=e^{\sigma}\tan {\tau+\phi\over 2},\quad \frac{\tilde {\sin}\phi}{\tilde {\sin}\tau}={\sin\phi\over \sin\tau}
$$
$$
\frac{\tilde {\cos}\phi}{\tilde {\sin}\tau}={\cos\phi\over \sin\tau}\cosh\sigma-{\cos \tau\over \sin \tau}\sinh\sigma,\quad 
\frac{\tilde {\cos}\tau}{\tilde {\sin}\phi}={\cos\tau\over\sin \phi}\cosh \sigma-{\cos\phi\over \sin\phi}\sinh\sigma
$$ 
$$
\frac{\tilde {\cos}\phi}{\tilde {\sin}\phi}={\cos\phi\over \sin\phi}\cosh\sigma-{\cos \tau\over \sin \phi}\sinh\sigma,\quad 
\frac{\tilde {\cos}\tau}{\tilde {\sin}\tau}={\cos\tau\over\sin \tau}\cosh \sigma-{\cos\phi\over \sin\tau}\sinh\sigma
$$ 
The formulae above define action of operator $e^{Q_{6,4}\sigma}$ in angle representation.

\subsubsection{The case $M^2\leq 0$}

In this case it is necessary to calculate proper functions of equation (\ref{IP}). But this equation coincide with (\ref{BUIT}) and calculation of the previous subsection may be repeated
with obvious exchanging energy on time, square of impulse on square of radius, square of mass on square of interval, $M^2\to L^2$. After this to constructed in such way solution it is necessary apply evolution operator $e^{Q_{6,4}\sigma}$ by the rules of the previous subsection. 
In the case under consideration symmetry of operator of square of mass and square of interval are the same $O(2,3)$. 

\subsubsection{The case $0\leq M^2$}

Let us transform equation (\ref{IH}) by tangent transformation $\cos^{-{3\over 2}}\phi A \cos^{{3\over 2}}\phi$ and represent  it in a form ( we also take into account proper value Kazimir operator if 4 dimensional compact group of rotation $K_2(O(4))=-n(n+2)$  
\begin{equation}
(\cos^2\phi\frac{\partial^2}{\partial \phi^2} -\cos\phi\sin\phi\frac{\partial}{\partial \phi}+((n+{1\over 2})(n+{3\over 2})\cos^2\phi-(t+{3\over 2})^2))\tilde F=0 \label{IN4}
\end{equation}
where $\tilde F=\cos^{-\rho-{3\over 2}}\phi e^{Q_{6,4}\sigma} F$.

\subsubsection{$t$ basis}

Operator of the time coordinate $\tilde x_4={\delta M x_4\over h}=iQ_{4,5}$ which equal (up to tangent transformations)
$$
-i{x_4\over L}=Q_{4,5}=-\sin\tau\sin\phi\frac{\partial}{\partial \tau}+\cos\phi\cos\tau\frac{\partial}{\partial \phi}
$$
It posses the following properties
$$
Q_{4,5} F({\sin\tau\over \cos\phi})=0, Q_{4,5} F({\sin\phi\over \cos\tau})=(1-({\sin\phi\over \cos\tau})^2 F_v
$$
In what follows ${\sin\tau\over \cos\phi}=u,{\sin\phi\over \cos\tau}=v$.
And thus proper function with proper value looks as
$$
e^{{-i\over 2}\ln {1+v \over v-1}\tilde x_4}F(u)
$$
where $x_4$ arbitrary real number.
Further equation (\ref{INN}) after substitution explicit expression for $K_2(O(4))$
may be rewritten in variables $u,v$ and after substitution explicit dependence from argument $v$ (found above) pass to ordinary differential equation with respect to argument $u$ ( these calculations are not complicate but not straight forward)
$$
(-u^2(1-u^2)F_{uu}+2uF_u+(t(t+3)-l(l+1)u^2-{u^2\over u^2-1}({x_4\over L})^2)F=0
$$
By exchange of variable $z\to {1\over u^2}$ pass to
$$
4z(1-z)F_{zz}-(10z-6)F_z+(t(t+3)-{l(l+1)\over z}+(\tilde x_4)^2{1\over 1-z})F=0
$$
By exchanging of unknown function $F\to z^{\alpha}(1-z)^{\beta}$ with $2\alpha=l,-(l+1),2\beta= 
\pm i {x_4\over L}$ the last equation pass to hupergeometrocal Gauss one. $a=\alpha+\beta-{t\over 2},b=\alpha+\beta+{t+3\over 2},c=2\alpha+{3\over 2}$. And finally for solution of (\ref{INN})equation we obtain
$$
F=e^{{-i\over 2}\ln {1+v \over v-1}\tilde x_4}z^{\alpha}(1-z)^{\beta}F_{2,1}(a,b,c,{1\over u^2})
$$

\subsubsection{$\rig{x}{}^2$ basis}

Really this problem was solved by (\ref{IN4}). Indeed $K_2(O(4))=\rig{x}{}^2+\rig{l}{}^2$
and we obtain for $\rig{x}{}^2=K_2(O(4))-\rig{l}{}^2=({h\over \delta M})^2[n(n+2)-l(l+1)$. 
The equation for proper function arises from (\ref{IN4}) After exchange of variable $ \phi\to {\pi\over 2}-\phi$ 
$$
(\frac{\partial^2}{\partial \phi^2} +{\cos\phi\over \sin\phi}\frac{\partial}{\partial \phi}+((n+{1\over 2})(n+{3\over 2})-{(t+{3\over 2})^2)\over \sin^2\phi})\tilde F=0 
$$
This is equation for usual spheric function with $l=n+{1\over 2},m=t+{3\over 2}$.

\section{Realization the alternative $O(2,4)$ space with $O(1,4)$ group of motion}

This case arises under condition $0 \leq L^2, \delta$ in (\ref{KAZ}). This case is different
from considered above only by exchanging by the places coordinate and impulses representations
with corresponding exchanging constants $L^2$ and $M^2$. Thus we will give belove only general formulae by help of which it is possible realize calculations independently. 

In the case under consideration \cite{LEZM} 7 generators are compact ones  
$$
{\delta\over h}l_{\alpha}={i\over 2}\sum \epsilon _{\alpha,\beta,\gamma}Q_{\beta,\gamma},\quad {\bar x_4\over L}=i\frac{\partial }{\partial \phi}
\quad {\delta L\over h} p_{\alpha}=i Q_{\alpha,4}
$$
are compact one and 8 are non compact ones
$$
-iI=\cos \phi(-\rho q^1_4+ \sum q^i_4 Q^{i,1})+
\sin \phi q^1_4 \frac{\partial}{\partial \phi}\equiv Q_{46}
$$
$$
-i{\delta L\over h} p_4=-\sin \phi(-\rho q^1_4+ \sum q^i_4 Q^{i,1})+
\cos \phi q^1_4 \frac{\partial}{\partial \phi}\equiv Q_{54}
$$
$$
-i{\bar x_{\alpha}\over L}=\cos \phi(-\rho q^1_{\alpha}+ \sum q^i_{\alpha} Q^{i,1})+
\sin \phi q^1_{\alpha} \frac{\partial}{\partial \phi}\equiv Q_{6\alpha}
$$
\begin{equation}
-i{\delta \over h} f_{\alpha}=-\sin \phi(-\rho q^1_{\alpha}+ \sum q^i_{\alpha} Q^{i,1})+
\cos \phi q^1_{\alpha} \frac{\partial}{\partial \phi}\equiv Q_{5\alpha} \label{REP1}
\end{equation}
where $Q_{ij},Q^{ij},1\leq i,j \leq 4$ generators if left (right) translations of compact four dimensional group of rotation.

\subsection{Impulse representation}

In connection with (\ref{REP1}) operator square of mass looks as
$$
({\delta L\over h})^2m^2=(Q^2_{\alpha,4}-Q^2_{\alpha,5}-Q^2_{5,4}+Q^2_{\alpha,\beta})
$$
This operator was calculated above (\ref{INN}) and its proper values and functions were found in subsection coordinate representation of the previous section.

\subsection{Coordinate representation}

All calculations of the corresponding subsections of the previous section are the same with only one difference - the cases $0 \leq M^2$ and $M^2\leq 0$ are changed by the places. And thus all necessary results may be rewritten from the same places of previous section.

\section{Matrix element of the free motion}

We can present here only integral form of this matrix element without its detail investigation of its properties.
$$
M(p,x)=\int \sin^2 \tau \sin \theta d\tau d\theta d\psi d\phi F(p,angles)\bar F(x,angles)   
$$
From classic solution it is known that there are existing domains forbidden for classical motion \cite{FM}. This means that above matrix element is going rapidly to zero in these domains.
And thus $M(p,x)\bar M(p,x)$ is not unity in the all alternative space. The finite value of alternative space possible lead to conclusion that that problem with divergences of field theory
in this space will be different than in usual Minkowski space-time one.

\section{Outlook}

In the present paper we have constructed mathematical formalism necessary to description events in an alternative world. This formalism is self- consistent and the remaining intriguing problem is to find a physical interpretation of it. We remind the reader
that all equations gauge and de-Sitter invariant field theory are written in non-physical angles or non-physical Minkowski space representation of conformal ($O(2,4)$) algebra. However, all the boundary conditions for them are stated in real alternative world. Of course, with the help of the formulae of the present paper it is possible to rewrite these equations for instance in $x$-representation. But these equation will be not local and involve not derivatives but finite differences, and it is not clear how to work with them. But all these are only technical problems.

From results of the present paper it follows that in alternative world energy and masses take natural values with coefficient ${h\over \delta L}=m_0$. Thus $m_0$ possible consider as a of mass quantum of mass in the alternative world. But, as it was mentioned above, this value is not related to Plank constant. Indeed $m_0^2={1\over ({L\over H})^2+{1\over M^2}}$ (for definitively we take $0 \leq M^2$ and $L^2\leq 0$). The same is true with respect to the distance measured in the units of $l_0^2={1\over ({M\over H})^2+{1\over L^2}}$. And thus in alternative world minimum and maximal observable are connected as ${m_0^2\over M^2}={l_0^2\over L^2}$ or, which is the same, $m_0^2={H^2\over L^2},l_0^2={H^2\over M^2}$.
From this expression it is possible estimate relation between macro constant from known from experiment (if any) values $m_0^2,l_0^2$.
 
It is possible interpret the physics of alternative world in extremal situations as follows. For observers at big distances ( but in the point of observation ) situation will be exactly the same as in the initial point of observation. But for us information that travels through great distances is distorted  by interacting with the matter on the way.
And thus in the point of observation we receive it only in probabilistic form. 

Of course situation will be much more clear after detail investigation of matrix element of free motion. In classical case this problem was solved in \cite{FM}, where was shown that there exists the domains forbidden for classical motion and, as was noted above, matrix element of free motion
must go to zero in this domains. 

In the present paper we investigated only the case of $O(2,4)$ alternative world. The author hopes in to give a similar solution in the cases of $O(1,5)$ and $O(3,3)$ algebras in the nearest time . In both these cases unity $I$ is compact generator and the space of representation with correct limit to Minkowski space-time will be discrete unitary representations of 6-dimensional algebras. In particular all alternative world will be constructed only in natural numbers. 

\section{Acknowledgments}

Author thanks J.Mostovoy for many fruitful discussions  and great help in preparation manuscript for publication and also CONNECUT for finance support.

\section{Appendix I}

Here we present formulae connected non physical Minkowski space-time with realization on two unite vectors
$$
x_4={\sin \tau\over \cos \tau-\cos \phi},\quad x_{\alpha}={q^1_{\alpha}\over \cos \tau-\cos \phi} 
$$
$$
\cos \phi={{x^2-1\over 2}\over \sqrt{({x^2-1\over 2})^2+x^2_4}},
\quad \cos \tau={{x^2+1\over 2}\over \sqrt{({x^2+1\over 2})^2+(\rig{x}{})^2}}
$$
The volume of representation space is finite with differential 
$$
d V=\sin^2 \tau \sin \theta d\tau d\theta d\psi d\phi={d^4 x\over [({x^2+1\over 2})^2+(\rig{x}{})^2]^2}
$$

\section{Appendix II}

In this Appendix we present explicit formulae in the case of $(1 + 1)$ alternative space to give possibility to the reader more visual understand calculations in the main text.
In this case conformal algebra in four dimensions $O(2.4)$ reduces up to $O(2,2)\simeq
O(1,2)\times O(1,2)$. Conserving in (\ref{REP}) indexes $i=1,4$ we rewrite them as  
$$
-iI=-{\rho\over 2}\cos \alpha ^++\sin \alpha ^+\frac{\partial}{\partial \alpha ^+}+
-{\rho\over 2}\cos \alpha ^-+\sin \alpha ^-\frac{\partial}{\partial \alpha ^-}
$$
$$
-i{\bar x_4\over L}={\rho\over 2}\sin \alpha ^++\cos \alpha ^+\frac{\partial}{\partial \alpha ^+}+{\rho\over 2}\sin \alpha ^-+\cos \alpha ^-\frac{\partial}{\partial \alpha ^-}-{h\over
\delta H}(\frac{\partial}{\partial \alpha ^+}+\frac{\partial}{\partial \alpha ^-})
$$
$$
-i{\bar x\over L}={h\over\delta H}({\rho\over 2}\sin \alpha ^++\cos \alpha ^+\frac{\partial}{\partial \alpha ^+}-{\rho\over 2}\sin \alpha ^--\cos \alpha ^-\frac{\partial}{\partial \alpha ^-})-(\frac{\partial}{\partial \alpha ^+}-\frac{\partial}{\partial \alpha ^-})
$$
$$
-i{\delta L\over h} p=-{\rho\over 2}\sin \alpha ^+-\cos \alpha ^+\frac{\partial}{\partial \alpha ^+}+{\rho\over 2}\sin \alpha ^-+\cos \alpha ^-\frac{\partial}{\partial \alpha ^-}
$$
$$
-i{\delta \over h} f=-{\rho\over 2}\cos \alpha ^++\sin \alpha ^+\frac{\partial}{\partial \alpha ^+}+{\rho\over 2}\cos \alpha ^--\sin \alpha ^-\frac{\partial}{\partial \alpha ^-}
$$
\begin{equation}
-i{\delta L\over h} p_4=\frac{\partial}{\partial \alpha ^+}+\frac{\partial}{\partial \alpha ^-} \label{REPA}
\end{equation}
where $\alpha ^{\pm}=\phi \pm \tau$.
The mass operator ( after absolutely manipulations ) takes the form
$$
({\delta L\over h})^2m^2=4\cos^{\rho+2}({\alpha ^+-\alpha ^-\over 2})\frac{\partial^2}{\partial \alpha ^-\partial \alpha ^+} \cos^{-\rho}({\alpha ^+-\alpha ^-\over 2})
$$
Three operators $p_4,p,f$ are commutative with $m^2$ and have common proper values and functions. In the first case $f_{p_4}=e^{-i\epsilon {\alpha ^++\alpha ^-\over 2}}F({\alpha ^+-\alpha ^-\over 2})$, where $\epsilon$ arbitrary natural number and proper value $p_4={h\over \delta L}\epsilon $. In the second one $f_p=({\tan ({\alpha ^+\over 2}-{\pi\over 4})\over \tan ({\alpha ^-\over 2}-{\pi\over 4}})^{-{ip\over 2}}F(\tan ({\alpha ^+\over 2}-{\pi\over 4})\tan ({\alpha^-\over 2}-{\pi\over 4})$, where $p^2$ arbitrary real number proper value of operator $p^2$. The third case is absolutely the sameone.
Substituting $f_{p_4}$ into mass operator we come to ordinary differential equation for $F\to \cos^{-\rho}F$ function in dimensionless variables
$$
{k(k+1)\over \cos^2\tau}F=F_{\tau,\tau}+\epsilon^2 F
$$
($m^2=k(k+1)$ as second order Kazimir operator of $O(1,2)$ algebra).
After introduction new argument $z=\sin^2\tau$ we pass from equation above to hypergeometric Gauss equation and obtain its general solution in a form
$$
f_{\epsilon}=\cos^{\rho+k+1}\tau e^{i\phi\epsilon}(A F_{2,1}({k+1+\epsilon\over 2},{k+1-\epsilon\over 2},{1\over 2},\sin^2\tau)+
$$
\begin{equation}
B \sin\tau F_{2,1}({k+2+\epsilon\over 2},{k+2-\epsilon\over 2},{3\over 2},\sin^2\tau))
\label{FFF1}
\end{equation}
In the second case substitute  $f_p$ into mass equation we obtain ordinary differential for
$F({\sin ({\alpha ^+\over 2}+{\alpha ^-\over 2}\over \cos ({\alpha ^+\over 2}-{\alpha ^-\over 2}})\equiv F(u)$
The last equation passes to Gauss hypergeometric function with finally result as
$$
f_p=({\tan ({\alpha ^+\over 2}-{\pi\over 4})\over \tan ({\alpha ^-\over 2}-{\pi\over 4}})^{-{ip\over 2}}\cos^{\rho+1}\tau (1-u^2)^{-ip\over 2} 
$$
\begin{equation}
(A F_{2.1}({-k+ip\over 2},{k+1+ip\over 2},{1\over 2},u^2)+B u F_{2.1}(({-k+1+ip\over 2},{k+2+ip\over 2},{3\over 2},u^2)\label{FFF2}
\end{equation}
In (un physical) Minkowski limit $\phi={x_4\over \delta L},\tau={x\over \delta L},\epsilon=p_4{\delta L\over h},p=p{\delta L\over h},m^2({\delta L\over h})^2=k(k+1)$
Two equalities with hypergeometric functions take place \cite{RG} (formula $9.121 (11-12)$
$$
F(k, l,{1\over 2},-{z^2\over 4kl})_{k\to \infty,l\to \infty}=\cos z, \quad
F(k, l,{3\over 2},-{z^2\over 4kl})_{k\to \infty,l\to \infty}={\sin z\over z}
$$
Substituting into (\ref{FFF1}) and (\ref{FFF2}) limiting Minkowski values and using the last equalities we obtain
$$
f_{\epsilon,m_1}=(ce^{i{x_4p_4\over h}}+de^{-i{x_4p_4\over h}})(a e^{i{x\sqrt {p^2_4-m_1^2}\over h}}+be^{-i{x\sqrt {p^2_4-m_1^2}\over h}}),
$$
$$
f_{p,m_2}=(ce^{i{x_4\sqrt {p^2+m_2^2}\over h}}+de^{-i{x_4\sqrt {p^2+m_2^2}\over h}})(a e^{i{xp\over h}}+be^{-i{xp\over h}})
$$
Matrix element 
$$
\int d x_4 d x f_{\epsilon,m_1}\bar f_{p,m_2}=c \delta (m_1-m_2) \delta(p_4\pm\sqrt {p^2+m^2})
$$

\end{document}